\documentclass[aps,pra,reprint,showpacs,floatfix,longbibliography]{revtex4-1}
\usepackage{graphicx, amsmath, amssymb}
\usepackage[caption=false]{subfig}
\allowdisplaybreaks[1] 
\begin{document}


\newcommand{\braket}[2]{{\left\langle #1 \middle| #2 \right\rangle}}
\newcommand{\bra}[1]{{\left\langle #1 \right|}}
\newcommand{\ket}[1]{{\left| #1 \right\rangle}}
\newcommand{\ketbra}[2]{{\left| #1 \middle\rangle \middle \langle #2 \right|}}


\title{Engineering the Success of Quantum Walk Search Using Weighted Graphs}

\author{Thomas G.~Wong}
	\email{twong@lu.lv}
	\affiliation{Faculty of Computing, University of Latvia, Rai\c{n}a bulv.~19, R\=\i ga, LV-1586, Latvia}

\author{Pascal Philipp}
	\email{pasc.philipp@gmail.com}
	\affiliation{National Laboratory for Scientific Computing, Petr\'{o}polis, CEP 25651-075, Rio de Janeiro, Brazil}

\begin{abstract}
	Continuous-time quantum walks are natural tools for spatial search, where one searches for a marked vertex in a graph. Sometimes, the structure of the graph causes the walker to get trapped, such that the probability of finding the marked vertex is limited. We give an example with two linked cliques, proving that the captive probability can be liberated by increasing the weights of the links. This allows the search to succeed with probability 1 without increasing the energy scaling of the algorithm. Further increasing the weights, however, slows the runtime, so the optimal search requires weights that are neither too weak nor too strong.
\end{abstract}


\maketitle


\section{Introduction}

Continuous-time quantum walks \cite{FG1998} are the quantum analogues of continuous-time classical random walks, or Markov chains, and they are the basis for a variety quantum algorithms. For example, they provide polynomial speedups over classical algorithms for solving the NAND tree problem \cite{FGG2008}, search \cite{CG2004}, and element distinctness \cite{Ambainis2003,Childs2010}. An exponential separation in black-box query complexity is even obtainable \cite{CCDFGS2003}.

In each of these algorithms, the edges of the graphs are unweighted (or equivalently, they all have weight $1$). In physical systems, however, this may not be the case. For example, continuous-time quantum walks underpin how photosynthetic systems transfer energy excitations in protein complexes \cite{Mohseni2008}, and the couplings in these structures may not all be equal. That is, nature seems to fine-tune the weights in the proteins to improve excitonic transport.

As such, it is of significant interest to investigate how quantum particles walk on weighted graphs, and whether the weights can be engineered to achieve desired goals. Some work on this has been done in the context of universal mixing \cite{CFHRTW2007} and quantum state transfer \cite{CDEL2004}. More recently, it was shown that by breaking time-reversal symmetry by manipulating the phases of the edges of a graph, one can achieve faster or more reliable transport \cite{Zimboras2013,Lu2016}.

In this paper, we show that weighted edges can be engineered to improve the success probability of an \emph{algorithm}. In particular, we improve a quantum walk's ability to solve spatial search \cite{CG2004}, where a quantum particle queries a Hamiltonian oracle \cite{Mochon2007} to find a ``marked'' vertex. If the graph is the complete graph, then it is equivalent to Grover's algorithm \cite{Grover1996,CG2004,AKR2005,Wong10}. For incomplete graphs, however, it is an open problem as to which graphs support fast quantum search in Grover's $O(\sqrt{N})$ time. Some graphs that have been analyzed include complete bipartite graphs \cite{Novo2015,Wong19}, hypercubes \cite{CG2004}, arbitrary dimensional square lattices \cite{CG2004}, balanced trees \cite{Philipp2016}, and Erd\"os-Renyi random graphs \cite{Chakraborty2016}.

\begin{figure}
\begin{center}
	\includegraphics{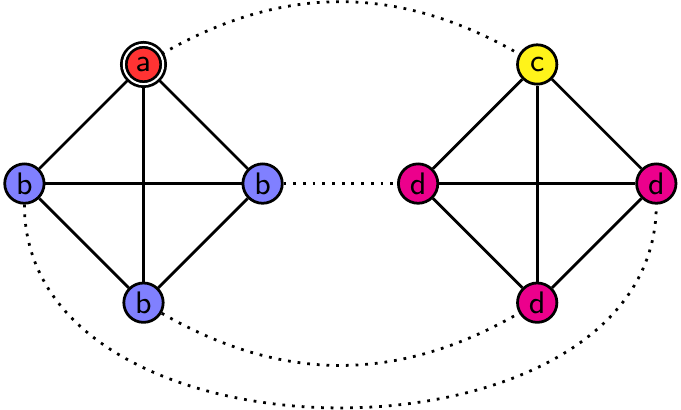}
	\caption{\label{fig:linkedcomplete} Two linked complete graphs, each of $M = 4$ vertices. Solid edges have weight $1$, and dotted edges have weight $w$. A vertex is marked, as indicated by a double circle. Identically evolving vertices are identically colored and labeled.}
\end{center}
\end{figure}

We focus on a new graph, which we construct from two complete graphs, each of $M$ vertices and fully connected within themselves with typical edges of weight $1$. Then we link these complete graphs together, pairing vertices between the cliques with edges of weight $w$. An example is illustrated in Fig.~\ref{fig:linkedcomplete}. Thus the total number of vertices is $N = 2M$, the number of edges of weight $1$ is $M(M-1)$ and the number of edges of weight $w$ is $M$.

This graph bears some resemblance to two previously studied (unweighted) graphs from \cite{Wong7}, the first of which has low connectivity but supports fast search, and the second of which has high connectivity but yields slow search. The first is also constructed from two complete graphs $K_M$, but rather than linking all vertices, only two vertices are joined together by a single edge. The second is constructed from $M+1$ complete graphs $K_M$, with each complete graph connected to the others through a single edge. This second graph is often called the ``simplex of complete graphs,'' and besides showing that high connectivity does not necessitate fast search \cite{Wong7}, it also shows a change in jumping rate depending on the arrangement of marked vertices \cite{Wong9}, and it demonstrates a faster continuous-time quantum walk search algorithm than the ``typical'' discrete-time one \cite{Wong11}.

Importantly for this work, the edges of the simplex of complete graphs can be weighted in such a way as to reduce the runtime of the quantum walk search from $O(N^{3/4})$ to nearly $\Theta(\sqrt{N})$ \cite{Wong16}. This seems to be the only prior work improving quantum walk search by engineering the weights of the graph. (There does exist a null result, however, where breaking time-reversal symmetry on the complete graph only changes the energy levels of the evolution without changing the runtime or success probability \cite{Wong14}.) In that work \cite{Wong16}, the success probability reached $1$ whether or not the graph was weighted. On the contrary, in the present work, the success probability is boosted from $1/2$ to $1$ using weighted edges.

In the next section, we formalize the problem of spatial search on the linked complete graphs. We show that the behavior of the algorithm depends on five different scalings for the weight $w$. In subsequent sections, we prove the behavior of the algorithm for the various weights, showing that as the weight increases, the success probability goes from $1/2$ to $1$. Further increasing the weights causes the runtime to worsen, however, while maintaining a success probability of $1$ through an inference. This eliminates the overall improvement from the probability boost. Thus there is a ``Goldilocks'' zone for the weights where the success probability is boosted to $1$, yet an overall runtime speedup is achieved.


\section{Quantum Walk Search}

The $N = 2M$ vertices of the linked complete graphs label computational basis states $\{ \ket{1}, \ket{2}, \dots, \ket{N} \}$. The system $\ket{\psi}$ begins in an equal superposition $\ket{s}$ over all the vertices:
\[ \ket{s} = \frac{1}{\sqrt{N}} \sum_{i=1}^N \ket{i}. \]
The system evolves in continuous-time by Schr\"odinger's equation with Hamiltonian \cite{CG2004}
\[ H = -\gamma L - \ketbra{a}{a}, \]
where $\gamma$ is a real parameter corresponding to the jumping rate (amplitude per time) of the walk, and $L$ is the graph Laplacian corresponding to the kinetic energy of a particle that is confined to discrete spatial locations. Together, $-\gamma L$ effects the quantum walk. In particular, $L = A - D$, where $A$ is the adjacency matrix of the graph ($A_{ij} = 1$ if vertices $i$ and $j$ are adjacent, and $0$ otherwise) and $D$ is the diagonal degree matrix ($D_{ii} = \text{deg}(i)$). The second term in the Hamiltonian $-\ketbra{a}{a}$ is the oracle, marking the vertex to search for. The goal is to find $\ket{a}$ in as little time as possible.

From the symmetry of this initial state and Hamiltonian, the system evolves such that there are only four types of vertices, as indicated in Fig.~\ref{fig:linkedcomplete}. Then the system evolves in a 4D subspace is spanned by
\begin{align*}
	\ket{a} &= \ket{\text{red}} \\
	\ket{b} &= \frac{1}{\sqrt{M-1}} \sum_{i \in \text{blue}} \ket{i} \\
	\ket{c} &= \ket{\text{yellow}} \\
	\ket{d} &= \frac{1}{\sqrt{M-1}} \sum_{i \in \text{magenta}} \ket{i}.
\end{align*}
In this $\{ \ket{a}, \ket{b}, \ket{c}, \ket{d} \}$ basis, the initial equal superposition state is
\[ \ket{s} = \frac{1}{\sqrt{2M}} \begin{pmatrix}
	1 \\
	\sqrt{M-1} \\
	1 \\
	\sqrt{M-1} \\
\end{pmatrix}. \]
The adjacency matrix is
\[ A = \begin{pmatrix}
	0 & \sqrt{M-1} & w & 0 \\
	\sqrt{M-1} & M-2 & 0 & w \\
	w & 0 & 0 & \sqrt{M-1} \\
	0 & w & \sqrt{M-1} & M-2 \\
\end{pmatrix}, \]
and the degree matrix is $D = (M+w-1)\mathbb{I}$. Since adding a multiple of the identity matrix to the Hamiltonian only constitutes a rezeroing of energy or multiplying by a global, unobservable phase \cite{CG2004,Wong19}, we can drop $D$ without changing the dynamics of the system. Then the search Hamiltonian is $H = -\gamma A - \ketbra{a}{a}$, which in the 4D basis is
\begin{equation}
	\label{eq:H}
	H = -\gamma \begin{pmatrix}
		\frac{1}{\gamma} & \sqrt{M-1} & w & 0 \\
		\sqrt{M-1} & M-2 & 0 & w \\
		w & 0 & 0 & \sqrt{M-1} \\
		0 & w & \sqrt{M-1} & M-2 \\
	\end{pmatrix}.
\end{equation}

Apart from the overall factor of $-\gamma$, this Hamiltonian has terms of various asymptotic scalings: constants, $\sqrt{M}$, and $M$. Then how the weight $w$ affects the evolution depends on its relation to these terms. In particular, there are five possible scalings, each with different dynamics, which we will work out shortly:
\begin{enumerate}
	\item	$w = o(\sqrt{M})$, or ``small'' weights.
	\item	$w = \Theta(\sqrt{M})$, or ``medium'' weights.
	\item	$w = \omega(\sqrt{M})$ and $w = o(M)$, or ``large'' weights.
	\item	$w = \Theta(M)$, or ``extra large (XL)'' weights.
	\item	$w = \omega(M)$, or ``extra extra large (XXL)'' weights.
\end{enumerate}
For example, when $w$ scales less than $\sqrt{M}$ (i.e., small weights), then we can drop it along with the constants because the most important terms are those that scale at least as big as $\sqrt{M}$ (this will be explicitly proven in the next section). For larger $w$, different terms contribute to the asymptotic evolution.

\begin{figure}
\begin{center}
	\subfloat[]{
		\label{fig:prob_time_M1000_w1-40}
		\includegraphics{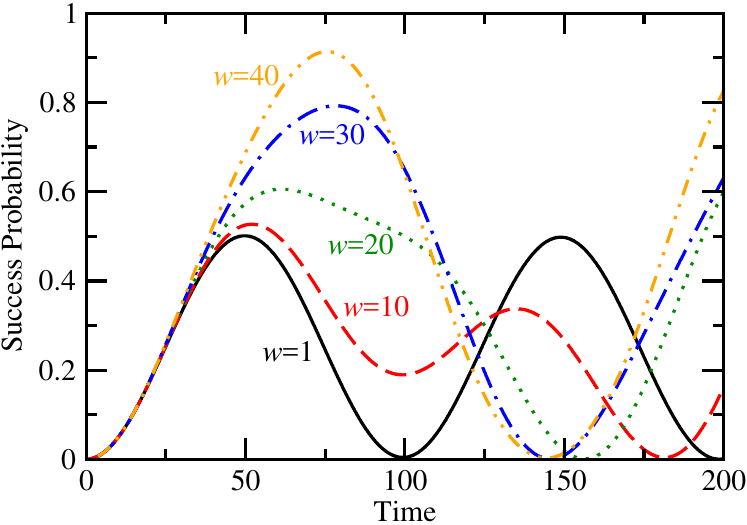}
	}

	\subfloat[]{
		\label{fig:prob_time_M1000_w100-20000}
		\includegraphics{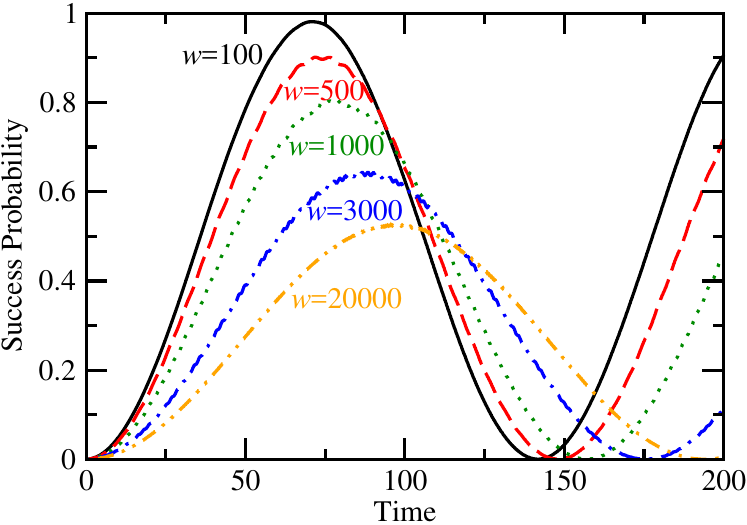}
	}
	\caption{\label{fig:prob_time_M1000} Success probability as a function of time for search on the linked complete graphs with $M = 1000$ and $\gamma = (M+w)/M(M+2w)$. (a) The solid black, dashed red, dotted green, dot-dashed blue, and dot-dot-dashed orange curves correspond to $w = 1$, $10$, $20$, $30$, and $40$, respectively. (b) The solid black, dashed red, dotted green, dot-dashed blue, and dot-dot-dashed orange curves correspond to $w = 100$, $500$, $1000$, $3000$, and $20000$, respectively.}
\end{center}
\end{figure}

To get a sense for the evolution, we plot the success probability $| \langle a | e^{-iHt} | s \rangle |^2$ as the system evolves with time in Fig.~\ref{fig:prob_time_M1000}. When $w = 1$, the graph is unweighted, and the success probability reaches $1/2$ at time $\pi\sqrt{M}/2 = \pi\sqrt{1000}/2 \approx 49.673$. As we will prove, the links are insignificant, so probability does not flow between the complete graphs. Thus the evolution is reduced to a single complete graph $K_M$ with total probability $1/2$. Note this behavior is identical to the complete graphs joined by a single edge in \cite{Wong7}, where half the probability is trapped in the unmarked clique.

Now as $w$ increases, Fig.~\ref{fig:prob_time_M1000} shows that the success probability increases, nearing $1$ at some point. So by manipulating the weights, we can engineer greater success of the search, liberating the trapped probability. If the weights are further increased, however, the success probability decreases back to $1/2$, and the time at which this maximum success probability is reached is twice that of the unweighted case. This is decrease is deceptive, however. With strong weights, the algorithm evolves to a superposition of $\ket{a}$ and $\ket{c}$. If one measures this superposition and gets vertex $\ket{c}$, then vertex $\ket{a}$ is immediately inferred as its neighbor in the other clique (see Fig.~\ref{fig:linkedcomplete}). With this inference, the true success probability actually remains at $1$. On the other hand, the slowdown is still genuine, enough so to eliminate the gains from the increased probability.

Thus there is an intermediate range of weights for which the search is improved, beyond which the increased success probability is offset by the slowdown. As we will show, this optimal weight corresponds to the large case, where $w$ scales greater than $\sqrt{M}$ but less than $M$.

In the next section, we work through the small weight case, showing that the weights are asymptotically negligible. This proof uses degenerate perturbation theory \cite{Wong5}. Afterward, we analyze the medium weight case, for which degenerate perturbation theory yields a transcendental equation for the runtime and success probability. Then we find the behavior of the large, extra large, and extra extra large weight cases, which can all be analyzed together, again using degenerate perturbation theory. All five cases are summarized in Table~\ref{table:summary}. Again, the large weight case is optimal, boosting the success probability to $1$ with minimal slowdown. Finally, we remark on the energy usage of the algorithm, that the overall scaling is unchanged with the large weights, so the improved probability by engineering the weights is energetically favorable.

\begin{table*}
	\caption{\label{table:summary} Summary of search on the linked complete graphs for various weights $w$. In the medium weight case, ``Transcendental'' means the runtime and success probability, and hence expected runtime, are given by a transcendental equation, and ``l.c.'' means a linear combination (i.e., superposition) of the states. In the XL and XXL cases, the ``success probability'' is the probability of measuring the final state in $\ket{a}$, but since $\ket{a}$ can be inferred from $\ket{c}$, the success probability is raised to $1$.}
	\begin{ruledtabular}
	\begin{tabular}{cccccc}
		Weight $w$ & Critical $\gamma$ & Runtime & Success Probability & Expected Runtime & Final State \\
		\colrule
		Small: $o(\sqrt{M})$ & $\frac{1}{M}$ & $\frac{\pi}{2} \sqrt{M}$ & $\frac{1}{2}$ & $\pi\sqrt{M}$ & $\frac{1}{\sqrt{2}} \left( e^{i\phi} \ket{a} + \ket{d} \right)$ \\
		Medium: $\Theta(\sqrt{M})$ & $\frac{1}{M+w}$ & Transcendental & Transcendental & Transcendental & l.c.~of $\ket{a}$, $\ket{b}$, and $\ket{d}$ \\
		Large: $\omega(\sqrt{M})$ and $o(M)$ & $\frac{M+w}{M(M+2w)}$ & $\frac{\pi}{\sqrt{2}} \sqrt{M}$ & $1$ & $\frac{\pi}{\sqrt{2}} \sqrt{M}$ & $\ket{a}$ \\
		Extra Large: $\Theta(M)$ & $\frac{M+w}{M(M+2w)}$ & $\frac{\pi\sqrt{M}\sqrt{(M+w)^2+w^2}}{\sqrt{2}(M+w)}$ & $\frac{(M+w)^2}{(M+w)^2 + w^2} \to 1$ & $\frac{\pi\sqrt{M}\sqrt{(M+w)^2+w^2}}{\sqrt{2}(M+w)}$ & $\frac{w}{\sqrt{(M+w)^2 + w^2}} \left( \frac{M+w}{w} \ket{a} + \ket{c} \right)$ \\
		Extra Extra Large: $\omega(M)$ & $\frac{M+w}{M(M+2w)}$ & $\pi \sqrt{M}$ & $\frac{1}{2} \to 1$ & $\pi \sqrt{M}$ & $\frac{1}{\sqrt{2}} \left( \ket{a} + \ket{c} \right)$ \\
	\end{tabular}
	\end{ruledtabular}
\end{table*}


\section{Small Weights}

\begin{figure}
\begin{center}
	\subfloat[]{
		\label{fig:diagram_H0_wS}
		\includegraphics{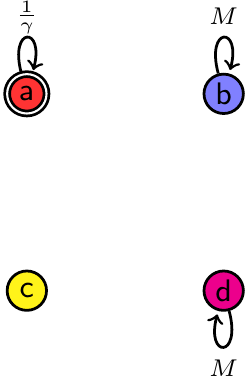}
	} \quad \quad \quad \quad
	\subfloat[]{
		\label{fig:diagram_H01_wS}
		\includegraphics{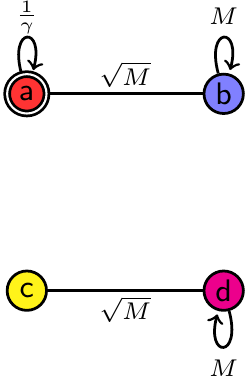}
	}
	\caption{\label{fig:diagram_wS} Apart from a factor of $-\gamma$, (a) the leading-order Hamiltonian with small weights, and (b) with a perturbation.}
\end{center}
\end{figure}

In this section, we consider the case of small weights, where $w$ scales less than $\sqrt{M}$. That is, $w = o(\sqrt{M})$. To find the evolution of the system, we want to find the eigenvectors and eigenvalues of the search Hamiltonian~\eqref{eq:H}. Unfortunately, directly finding the eigensystem of \eqref{eq:H} is arduous. So we instead approximate it for large $M$ using degenerate perturbation theory \cite{Wong5}. To leading order, the search Hamiltonian \eqref{eq:H} is
\[ H^{(0)} = -\gamma \begin{pmatrix}
	\frac{1}{\gamma} & 0 & 0 & 0 \\
	0 & M & 0 & 0 \\
	0 & 0 & 0 & 0 \\
	0 & 0 & 0 & M \\
\end{pmatrix}. \]
This is diagrammatically \cite{Wong8} represented in Fig.~\ref{fig:diagram_H0_wS}. So the eigenvectors of $H^{(0)}$ are $\ket{a}$, $\ket{b}$, $\ket{c}$, and $\ket{d}$ with corresponding eigenvalues $-1$, $-\gamma M$, $0$, and $-\gamma M$. Note that $\ket{b}$ and $\ket{d}$ are degenerate, and the initial state $\ket{s}$ is approximately $(\ket{b} + \ket{d})/\sqrt{2}$. Then for the system to evolve to $\ket{a}$, we want $\ket{a}$ to also be degenerate with $\ket{b}$ and $\ket{d}$. This requires setting $\gamma$ equal to its ``critical value'' of
\[ \gamma_c = \frac{1}{M}. \]
If $\gamma$ is chosen away from $\gamma_c$, then the initial state of the system is an asymptotic eigenvector of $H$, and the system only evolves by a trivial global phase \cite{CG2004,Wong5}.

The perturbation $H^{(1)}$ restores terms that scale as $\sqrt{M}$, so the perturbed Hamiltonian is
\[ H^{(0)} + H^{(1)} = -\gamma \begin{pmatrix}
	\frac{1}{\gamma} & \sqrt{M} & 0 & 0 \\
	\sqrt{M} & M & 0 & 0 \\
	0 & 0 & 0 & \sqrt{M} \\
	0 & 0 & \sqrt{M} & M \\
\end{pmatrix}. \]
This is diagrammatically \cite{Wong8} shown in Fig.~\ref{fig:diagram_H01_wS}, and it indicates that probability may possibly flow from $\ket{b}$ to $\ket{a}$ and from $\ket{d}$ to $\ket{c}$. So we already see that the success probability can be at most $1/2$. With the perturbation, linear combinations of $\ket{a}$, $\ket{b}$, and $\ket{d}$
\[ \alpha_a \ket{a} + \alpha_b \ket{b} + \alpha_d \ket{d} \]
become eigenvectors of the perturbed Hamiltonian, where the coefficients are found by solving
\[ \begin{pmatrix}
	H_{aa} & H_{ab} & H_{ad} \\
	H_{ba} & H_{bb} & H_{bd} \\
	H_{da} & H_{db} & H_{dd} \\
\end{pmatrix} \begin{pmatrix}
	\alpha_a \\
	\alpha_b \\
	\alpha_d \\
\end{pmatrix} = E \begin{pmatrix}
	\alpha_a \\
	\alpha_b \\
	\alpha_d \\
\end{pmatrix}, \]
where $H_{ab} = \bra{a} H^{(0)}+H^{(1)} \ket{b}$, etc. Note that $\ket{c}$ remains an approximate eigenvector of the perturbed system, but it is not relevant for the evolution of the system. Evaluating these matrix components with $\gamma = \gamma_c = 1/M$, we get
\[ \begin{pmatrix}
	-1 & \frac{-1}{\sqrt{M}} & 0 \\
	\frac{-1}{\sqrt{M}} & -1 & 0 \\
	0 & 0 & -1 \\
\end{pmatrix} \begin{pmatrix}
	\alpha_a \\
	\alpha_b \\
	\alpha_d \\
\end{pmatrix} = E \begin{pmatrix}
	\alpha_a \\
	\alpha_b \\
	\alpha_d \\
\end{pmatrix}. \]
Solving this eigenvalue problem, the perturbed eigenvectors and eigenvalues are
\begin{gather*}
	\ket{\psi_-} = \frac{1}{\sqrt{2}} \left( \ket{b} + \ket{a} \right), \quad E_- = -1 - \frac{1}{\sqrt{M}} \\
	\ket{\psi_+} = \frac{1}{\sqrt{2}} \left( \ket{b} - \ket{a} \right), \quad E_+ = -1 + \frac{1}{\sqrt{M}} \\
	\ket{d}, \quad E_{-1} = -1.
\end{gather*}
Recall that the initial equal superposition state $\ket{s} \approx (\ket{b} + \ket{d}) / \sqrt{2}$, but the $\ket{d}$ component does not appreciably evolve since it is an approximate eigenvector of $H$ (with eigenvalue $-1$). So we only care about how $\ket{b}$ evolves. Since $(\ket{b} \pm \ket{a})/\sqrt{2}$ are approximate eigenvectors of $H$, we get that $\ket{b}$ evolves to $\ket{a}$ (up to a phase) in time $\pi/\Delta E$, where $\Delta E$ is the energy gap between the two eigenvectors \cite{CG2004,Wong10}. Thus the system evolves from $\ket{s} \approx (\ket{b} + \ket{d}) / \sqrt{2}$ to $(e^{i\phi} \ket{a} + \ket{d}) / \sqrt{2}$, for some phase $\phi$ and up to a global phase, which corresponds to a success probability of
\[ p_* = \frac{1}{2} \]
at time
\[ t_* = \frac{\pi}{\Delta E} = \frac{\pi}{2} \sqrt{M}. \]
This proves that the system asymptotically evolves as if the two cliques were disconnected. In the clique containing the marked vertex, the system roughly evolves from $\ket{b}$ to $\ket{a}$ in time $\pi\sqrt{M}/2$, which is the expected runtime for a complete graph of $M$ vertices \cite{CG2004,Wong10}. The other clique, which is unmarked, roughly stays in $\ket{d}$ up to a global phase, since $\ket{d}$ approximates the uniform superposition over the vertices of the clique, which is an eigenvector of the quantum walk \cite{Wong14}.

Since the success probability of a single run of the algorithm is $1/2$, we expect to classically repeat the algorithm twice in order to find the marked vertex. Thus the expected runtime is twice that of a single runtime, i.e., $\pi\sqrt{M}$. This result is summarized in Table~\ref{table:summary}.


\section{Medium Weights}

Now we consider the medium weight case, where $w$ scales as $\sqrt{M}$. That is, $w = \Theta(\sqrt{M})$. If we naively employ degenerate perturbation theory \cite{Wong5}, we might use the same leading-order search Hamiltonian as in the small weight case, which was visualized in Fig.~\ref{fig:diagram_H0_wS}. Then it seems as though the critical value of $\gamma$ should be $1/M$ so that $\ket{a}$, $\ket{b}$, and $\ket{d}$ are degenerate to leading order.

\begin{table}
	\caption{\label{table:edges} The types of edges in the linked complete graphs, with their respective weight and the number of them.}
	\begin{ruledtabular}
	\begin{tabular}{ccc}
		Connection & Weight & Number of Edges \\
		\colrule
		$a \sim b$ & $1$ & $M-1$ \\
		$a \sim c$ & $w$ & $1$ \\
		$b \sim b$ & $1$ & $\frac{(M-1)(M-2)}{2}$ \\
		$b \sim d$ & $w$ & $M-1$ \\
		$c \sim d$ & $1$ & $M-1$ \\
		$d \sim d$ & $1$ & $\frac{(M-1)(M-2)}{2}$ \\
	\end{tabular}
	\end{ruledtabular}
\end{table}

\begin{figure}
\begin{center}
	\subfloat[]{
		\label{fig:diagram_H0_wM}
		\includegraphics{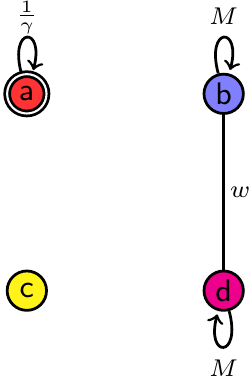}
	} \quad \quad \quad \quad
	\subfloat[]{
		\label{fig:diagram_H01_wM}
		\includegraphics{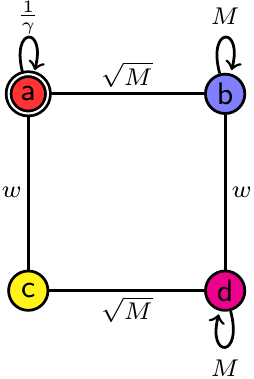}
	}
	\caption{\label{fig:diagram_wL} Apart from a factor of $-\gamma$, (a) the leading-order Hamiltonian with medium weights, and (b) with a perturbation.}
\end{center}
\end{figure}

This initial attempt, however, neglects some crucial edges \cite{Wong16}. Counting the number of each type of edge, shown in Table~\ref{table:edges}, we see that $b \sim b$ and $d \sim d$ dominate for large $M$. But they are already included in $H^{(0)}$, manifesting themselves as the diagonal $M$ terms. The next most significant type of edge is $b \sim d$, so we add it to $H^{(0)}$, yielding
\[ H^{(0)} = -\gamma \begin{pmatrix}
	\frac{1}{\gamma} & 0 & 0 & 0 \\
	0 & M & 0 & w \\
	0 & 0 & 0 & 0 \\
	0 & w & 0 & M \\
\end{pmatrix}. \]
This is depicted in Fig.~\ref{fig:diagram_H0_wM}. Note that $a \sim b$ and $c \sim d$ are less significant than $b \sim d$, even though they have the same number of edges, because their weight of $1$ is less than the weight $w = \Theta(\sqrt{M})$. The eigenvectors and eigenvalues of the adjusted leading-order Hamiltonian $H^{(0)}$ are
\begin{gather*}
	\ket{a}, \quad -1 \\
	\ket{\sigma} = \frac{1}{\sqrt{2}} \left( \ket{b} + \ket{d} \right), \quad -\gamma(M+w) \\
	\ket{\delta} = \frac{1}{\sqrt{2}} \left( -\ket{b} + \ket{d} \right), \quad -\gamma(M-w) \\
	\ket{c}, \quad 0.
\end{gather*}
Setting the first two eigenvalues equal to each other, we get the critical $\gamma$:
\[ \gamma_c = \frac{1}{M+w}. \]
Taylor expanding this, we get
\[ \frac{1}{M+w} = \frac{1}{M} - \frac{w}{M^2} + O\left( \frac{w^2}{M^3} \right), \]
so this corrects the naive critical value of $\gamma$ of $1/M$ by a term $-w/M^2$.

\begin{figure}
\begin{center}
	\includegraphics{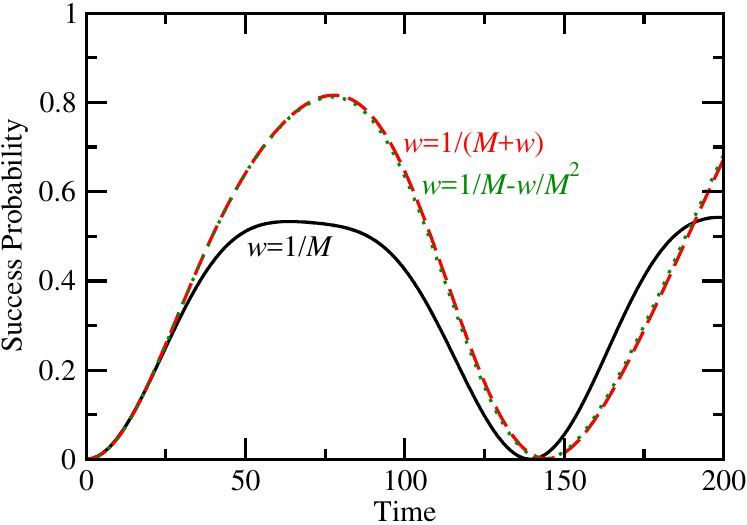}
	\caption{\label{fig:prob_time_M1000_wsqrtM_g} Success probability as a function of time for search on the linked complete graphs with $M = 1000$ and $w = \sqrt{M}$. The solid black, dashed red, and dotted green curves corresponds to $\gamma = 1/M$, $\gamma = 1/(M+w)$$, and \gamma = 1/M - w/M^2$, respectively.}
\end{center}
\end{figure}

We now prove that this correction $-w/M^2$ is significant. Using the argument from Section VI of \cite{Wong16}, say $\gamma = \gamma_c + \epsilon$. Then $\bra{b} H^{(0)} + H^{(1)} \ket{b}$ contributes a term $\epsilon M$ to the perturbative calculation, and this is the leading-order term in $\epsilon$. We want it to scale less than the energy gap of $1/\sqrt{M}$ so that the runtime and success probability are asymptotically correct. That is, $\epsilon M = o(1/\sqrt{M})$, which implies that $\epsilon = o(1/M^{3/2})$. This specifies the precision to which $\gamma_c$ must be known. Since for medium weights $w = \Theta(\sqrt{M})$, the correction $-w/M^2 = \Theta(1/M^{3/2})$, which is significant enough to affect the algorithm, and so it must be included. This is confirmed in Fig.~\ref{fig:prob_time_M1000_wsqrtM_g}, where using $\gamma = 1/M$ yields a worse algorithm than $\gamma = 1/(M+w)$ or $\gamma = 1/M - w/M^2$.

Note that the significance of $-w/M^2$ depends on the weight $w$. For an unweighted graph (i.e., with $w = 1$), the correction would scale as $1/M^2$, as expected from Eq.~(7) of \cite{Wong20}. With small weights $w = o(\sqrt{M})$, the correction $-w/M^2 = o(1/M^{3/2})$ is small enough to be dropped. It is only with medium or greater weights that the correction becomes important.

Returning to the perturbative calculation, with $\gamma_c = 1/(M+w)$, the leading-order eigenstate $\ket{\delta} = (-\ket{b} + \ket{d})/\sqrt{2}$ has eigenvalue
\[ -\gamma(M-w) = -\frac{M-w}{M+w} = -1 + \frac{2w}{M-w}. \]
When $w = \Theta(\sqrt{M})$, the last term is $\Theta(1/\sqrt{M})$, which is significant enough to affect the perturbative calculation since it scales as the energy gap. So $\ket{\delta}$ is also approximately degenerate with $\ket{a}$ and $\ket{\sigma}$.

With the perturbation $H^{(1)}$, which restores terms of $\Theta(\sqrt{M})$, we get
\[ H^{(0)} + H^{(1)} =  -\gamma \begin{pmatrix}
	\frac{1}{\gamma} & \sqrt{M} & w & 0 \\
	\sqrt{M} & M & 0 & w \\
	w & 0 & 0 & \sqrt{M} \\
	0 & w & \sqrt{M} & M \\
\end{pmatrix}, \]
as depicted in Fig.~\ref{fig:diagram_H01_wM}. This causes linear combinations $\alpha_a \ket{a} + \alpha_\sigma \ket{\sigma} + \alpha_\delta \ket{\delta}$ to become eigenstates of the perturbed system, where
\[ \begin{pmatrix}
	H_{aa} & H_{a\sigma} & H_{a\delta} \\
	H_{\sigma a} & H_{\sigma\sigma} & H_{\sigma\delta} \\
	H_{\delta a} & H_{\delta\sigma} & H_{\delta\delta} \\
\end{pmatrix} \begin{pmatrix}
	\alpha_a \\
	\alpha_\sigma \\
	\alpha_\delta \\
\end{pmatrix} = E \begin{pmatrix}
	\alpha_a \\
	\alpha_\sigma \\
	\alpha_\delta \\
\end{pmatrix}, \]
where $H_{a\sigma} = \bra{a} H^{(0)}+H^{(1)} \ket{\sigma}$, etc. Evaluating these matrix components with $\gamma = \gamma_c = 1/(M+w)$, we get
\[ \begin{pmatrix}
	-1 & \frac{-\sqrt{M}}{\sqrt{2}(M+w)} & \frac{\sqrt{M}}{\sqrt{2}(M+w)} \\
	\frac{-\sqrt{M}}{\sqrt{2}(M+w)} & -1 & 0 \\
	\frac{\sqrt{M}}{\sqrt{2}(M+w)} & 0 & -1 + \frac{2w}{M+w} \\
\end{pmatrix} \begin{pmatrix}
	\alpha_a \\
	\alpha_\sigma \\
	\alpha_\delta \\
\end{pmatrix} = E \begin{pmatrix}
	\alpha_a \\
	\alpha_\sigma \\
	\alpha_\delta \\
\end{pmatrix} \!. \]
Solving this takes some work. Let us call the $3 \times 3$ matrix $H'$. Since adding a multiple of the identity matrix only constitues a rezeroing of energy, or multiplying by a global phase, we drop the $-1$'s on the diagonal. Then factoring out $1/(M+w)$, we get
\[ H' = \frac{1}{M+w} \begin{pmatrix}
	0 & -\sqrt{\frac{M}{2}} & \sqrt{\frac{M}{2}} \\
	-\sqrt{\frac{M}{2}} & 0 & 0 \\
	\sqrt{\frac{M}{2}} & 0 & 2w \\
\end{pmatrix}. \]
Now let $w = k\sqrt{M}$. Then factoring out $\sqrt{M/2}$, we get
\[ H' = \frac{\sqrt{M}}{\sqrt{2}(M+w)} \underbrace{\begin{pmatrix}
	0 & -1 & 1 \\
	-1 & 0 & 0 \\
	1 & 0 & 2\sqrt{2} k
\end{pmatrix}}_{H''}. \]
We can find the eigenvalues and eigenvectors of $H''$, which only has a single variable $k$. Say $\vec{v}$ is an eigenvector of $H''$ with eigenvalue $\lambda$, \textit{i.e.}, $H'' \vec{v} = \lambda \vec{v}$. Then it is also an eigenvector of $H'$ with eigenvalue $(\sqrt{M} / \sqrt{2}(M+w)) \lambda$. Note that $H'$ and $H''$ are both in the $\{ \ket{a}, \ket{\sigma}, \ket{\delta} \}$ basis. 

The characteristic equation of $H''$ is
\[ \lambda^3 - 2\sqrt{2}k \lambda^2 - 2 \lambda + 2\sqrt{2}k = 0. \]
Solving this yields
\begin{align*}
	\lambda_1 &= \frac{2\sqrt{2}}{3} \left[ k + \sqrt{4k^2 + 3} \cos\left( \frac{\theta}{3} \right) \right] \\
	\lambda_2 &= \frac{2\sqrt{2}k}{3} - \frac{\sqrt{2}\sqrt{4k^2 + 3}}{3} \left[ \cos\left( \frac{\theta}{3} \right) + \sqrt{3} \sin\left( \frac{\theta}{3} \right) \right] \\
	\lambda_3 &= \frac{2\sqrt{2}k}{3} - \frac{\sqrt{2}\sqrt{4k^2 + 3}}{3} \left[ \cos\left( \frac{\theta}{3} \right) - \sqrt{3} \sin\left( \frac{\theta}{3} \right) \right],
\end{align*}
where
\[ \cos\theta = \frac{k (16k^2 - 9)}{2(4k^2+3)^{3/2}}, \quad \sin\theta = \frac{3\sqrt{3}\sqrt{32k^4 + 13k^3 + 4}}{2(4k^2+3)^{3/2}}. \]

Now let us find the corresponding eigenvectors $\ket{\psi_i} = (a \ \sigma \ \delta)^\intercal$, which satisfy $H'' \ket{\psi_i} = \lambda\ket{\psi_i}$:
\[ \begin{pmatrix}
	0 & -1 & 1 \\
	-1 & 0 & 0 \\
	1 & 0 & 2\sqrt{2} k \\
\end{pmatrix} \begin{pmatrix}
	a \\
	\sigma \\
	\delta \\
\end{pmatrix} = \lambda \begin{pmatrix}
	a \\
	\sigma \\
	\delta \\
\end{pmatrix}. \]
The second line yields
\[ -a = \lambda \sigma \quad \Rightarrow \quad \sigma = \frac{-a}{\lambda}, \]
and the first line, with substitution of the second line, yields
\[ -\sigma + \delta = \lambda a \quad \Rightarrow \quad \frac{a}{\lambda} + \delta = \lambda a \quad \Rightarrow \quad \delta = \left( \lambda - \frac{1}{\lambda} \right) a. \]
So the (unnormalized) eigenvectors of $H''$ are
\[ \ket{\psi_i} = \begin{pmatrix}
	1 \\
	\frac{-1}{\lambda_i} \\
	\lambda_i - \frac{1}{\lambda_i} \\
\end{pmatrix}. \]

Since the system starts in $\ket{s} \approx \ket{\sigma}$, want to find the superposition of eigenvectors that equals $\ket{\sigma}$. That is, we want to find $\alpha_1$, $\alpha_2$, and $\alpha_3$ such that
\[ \ket{\sigma} = \alpha_1\ket{\psi_1} + \alpha_2 \ket{\psi_2} + \alpha_3 \ket{\psi_3}. \]
In the $\{ \ket{a}, \ket{\sigma}, \ket{\delta} \}$ basis, $\ket{\sigma} = (0 \ 1 \ 0)^\intercal$, so when plugging in for the eigenvectors $\ket{\psi_i}$, we get three equations
\begin{gather*}
	\alpha_1 + \alpha_2 + \alpha_3 = 0 \\
	-\frac{\alpha_1}{\lambda_1} - \frac{\alpha_2}{\lambda_2} - \frac{\alpha_3}{\lambda_3} = 1 \\
	\alpha_1 \left( \lambda_1 - \frac{1}{\lambda_1} \right) + \alpha_2 \left( \lambda_2 - \frac{1}{\lambda_2} \right) + \alpha_3 \left( \lambda_3 - \frac{1}{\lambda_3} \right) = 0.
\end{gather*}
Using the second equation, we can simplify the third equation to be
\[ \alpha_1 \lambda_1 + \alpha_2 \lambda_2 + \alpha_3 \lambda_3 = -1. \]
Solving this system of three equations yields
\begin{align*}
	\alpha_1 &= - \frac{\lambda_1 + \lambda_1 \lambda_2 \lambda_3}{(\lambda_1 - \lambda_2)(\lambda_1 - \lambda_3)} \\
	\alpha_2 &= \frac{\lambda_2 + \lambda_1 \lambda_2 \lambda_3}{(\lambda_1 - \lambda_2)(\lambda_2 - \lambda_3)} \\
	\alpha_3 &= - \frac{\lambda_3 + \lambda_1 \lambda_2 \lambda_3}{(\lambda_1 - \lambda_3)(\lambda_2 - \lambda_3)}.
\end{align*}
Then the state of the system at time $t$ is
\[ e^{-iHt} \ket{\sigma} = \alpha_1 e^{-i \lambda_1 t} \ket{\psi_1} + \alpha_2 e^{-i \lambda_2 t} \ket{\psi_2} + \alpha_3 e^{-i \lambda_3 t} \ket{\psi_3}. \]
Taking the inner product with $\bra{a}$ and using $\braket{a}{\psi_i} = 1$, the success amplitude is
\[ \bra{a} e^{-iHt} \ket{\sigma} = \alpha_1 e^{-i \lambda_1 t} + \alpha_2 e^{-i \lambda_2 t} + \alpha_3 e^{-i \lambda_3 t}. \]
Taking the norm square, the success probability is
\begin{align}
	p(t)
		&= \alpha_1^2 + \alpha_2^2 + \alpha_3^2 + 2\alpha_1\alpha_2 \cos[(\lambda_1-\lambda_2)t] \label{eq:medium_prob} \\
		&\quad+ 2\alpha_1\alpha_3 \cos[(\lambda_1-\lambda_3)t] + 2\alpha_2\alpha_3 \cos[(\lambda_2-\lambda_3)t]. \notag
\end{align}

\begin{figure}
\begin{center}
	\includegraphics{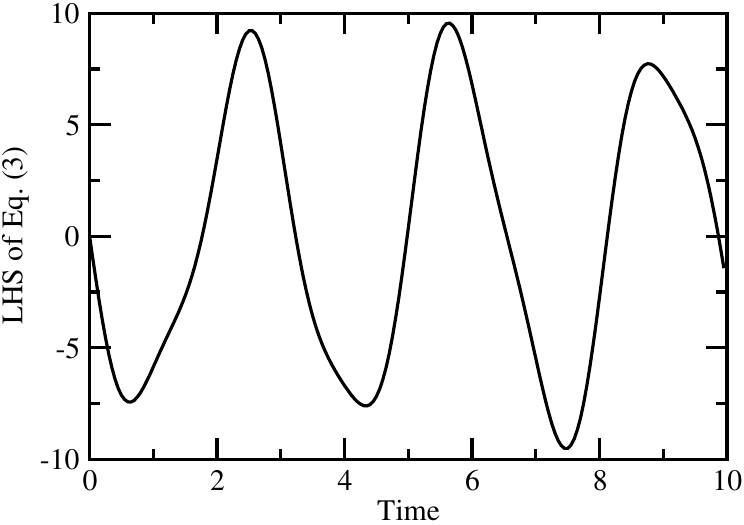}
	\caption{\label{fig:trans_M1000_wsqrtM} The left-hand side of the transcendental equation \eqref{eq:medium_runtime} with $M = 1000$ and $w = \sqrt{M}$ (i.e., $k = 1$).}
\end{center}
\end{figure}

To find the runtime, we find the first maximum in success probability by solving $dp/dt = 0$. The derivative of the success probability $p(t)$ is
\begin{align*}
	\frac{dp}{dt} 
		&= -2\alpha_1\alpha_2(\lambda_1-\lambda_2)\sin[(\lambda_1-\lambda_2)t] \\[-0.07in]
		&\quad -2\alpha_1\alpha_3(\lambda_1-\lambda_3)\sin[(\lambda_1-\lambda_3)t] \\
		&\quad -2\alpha_2\alpha_3(\lambda_2-\lambda_3)\sin[(\lambda_2-\lambda_3)t].
\end{align*}
Plugging in for $\alpha_1$, $\alpha_2$, and $\alpha_3$,
\begin{align*}
	\frac{dp}{dt}
		&= \frac{2 (\lambda_1 + \lambda_1\lambda_2\lambda_3) (\lambda_2 + \lambda_1\lambda_2\lambda_3) \sin[(\lambda_1-\lambda_2)t]}{(\lambda_1-\lambda_2)(\lambda_1-\lambda_3)(\lambda_2-\lambda_3)} \\
		&\quad- \frac{2 (\lambda_1 + \lambda_1\lambda_2\lambda_3) (\lambda_3 + \lambda_1\lambda_2\lambda_3) \sin[(\lambda_1-\lambda_3)t]}{(\lambda_1-\lambda_2)(\lambda_1-\lambda_3)(\lambda_2-\lambda_3)} \\
		&\quad+ \frac{2 (\lambda_2 + \lambda_1\lambda_2\lambda_3) (\lambda_3 + \lambda_1\lambda_2\lambda_3) \sin[(\lambda_2-\lambda_3)t]}{(\lambda_1-\lambda_2)(\lambda_1-\lambda_3)(\lambda_2-\lambda_3)}.
\end{align*}
Setting this equal to zero and simplifying,
\begin{align}
	&(\lambda_1 + \lambda_1\lambda_2\lambda_3) (\lambda_2 + \lambda_1\lambda_2\lambda_3) \sin[(\lambda_1-\lambda_2)t] \notag \\
	&\quad- (\lambda_1 + \lambda_1\lambda_2\lambda_3) (\lambda_3 + \lambda_1\lambda_2\lambda_3) \sin[(\lambda_1-\lambda_3)t] \label{eq:medium_runtime} \\
	&\quad+ (\lambda_2 + \lambda_1\lambda_2\lambda_3) (\lambda_3 + \lambda_1\lambda_2\lambda_3) \sin[(\lambda_2-\lambda_3)t] = 0. \notag
\end{align}
We get a transcendental equation. For example, we plot the left-hand side of it in Fig.~\ref{fig:trans_M1000_wsqrtM} with $M = 1000$ and $w = \sqrt{M}$ (i.e., $k = 1$). We are interested in the first nonzero root of this, which is around $t = 1.766$. But this time corresponds to $H''$, not $H'$. So to get the actual runtime $t_*$, we need to multiply it by a rescaling factor:
\[ t_* = \frac{\sqrt{2}(M+w)}{\sqrt{M}} \cdot \text{[First nonzero root of \eqref{eq:medium_runtime}]}. \]
Continuing with our example from Fig.~\ref{fig:trans_M1000_wsqrtM}, the runtime is $\sqrt{2}(1000 + \sqrt{1000})(1.766)/\sqrt{1000} \approx 81.45$. This has fair agreement with Fig.~\ref{fig:prob_time_M1000_wsqrtM_g}; it is slightly large, but as we will see, the discrepancy is negligible.

To get the success probability, we plug the first nonzero root of \eqref{eq:medium_runtime} (without rescaling it) into \eqref{eq:medium_prob}. Again with our example from Fig.~\ref{fig:trans_M1000_wsqrtM}, we plug $t = 1.766$ into \eqref{eq:medium_prob} and get a success probability of $p_* = 0.82$, which has great agreement with Fig.~\ref{fig:prob_time_M1000_wsqrtM_g}.

\begin{figure}
\begin{center}
	\subfloat[]{
		\label{fig:prob_k_M1000} 
		\includegraphics{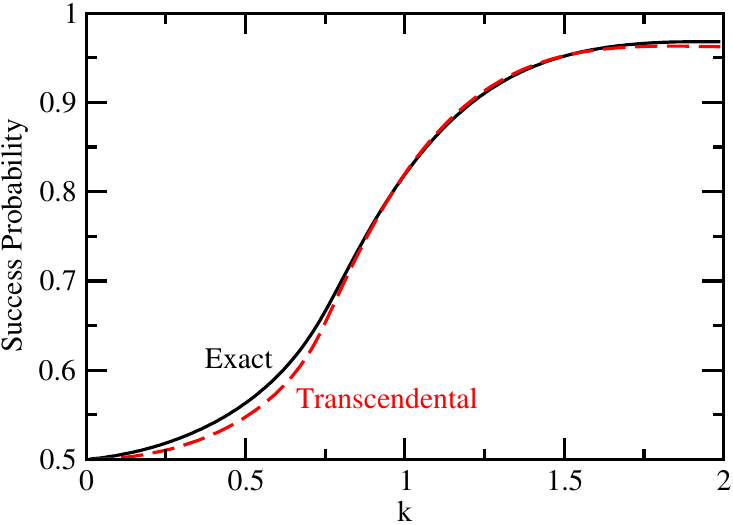}
	}

	\subfloat[]{
		\label{fig:time_k_M1000} 
		\includegraphics{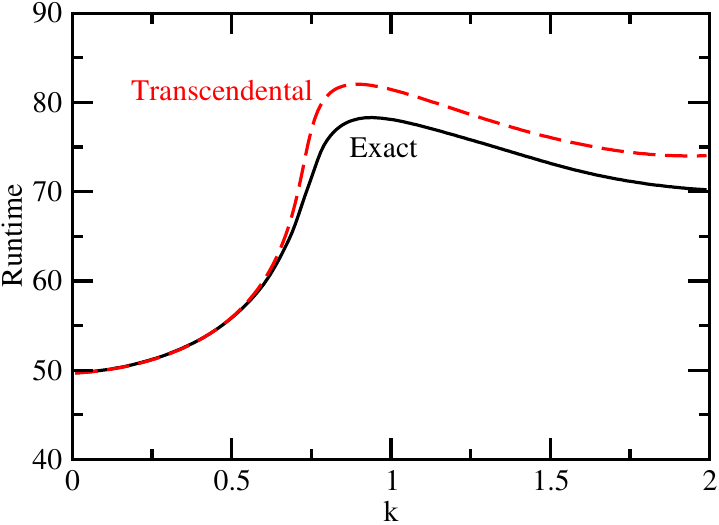}
	}
	\caption{(a) Success probability and (b) runtime of the search algorithm as a function of $k$, with $M = 1000$, $w = k\sqrt{M}$, and $\gamma = (M+w)/M(M+2w)$. The solid black curve comes from numerically evolving the system from $\ket{s}$ using the exact search Hamiltonian \eqref{eq:H}, and the dashed red curve comes from transcendental equations \eqref{eq:medium_prob} and \eqref{eq:medium_runtime}.}
\end{center}
\end{figure}

Thus for the medium weight case, our perturbative approximation for the eigenvectors and eigenvalues of the search Hamiltonian yields transcendental equations for the runtime and success probability. This is summarized in Table~\ref{table:summary}. Since medium weights are when the success probability increases from $1/2$ (for small weights) to $1$ (for large weights, proved next), we plot this transition in Fig.~\ref{fig:prob_k_M1000}. We see that the transcendental equations are in fairly close agreement with the success probability obtained by numerically evolving the system from $\ket{s}$ using the exact $4 \times 4$ search Hamiltonian \eqref{eq:H}.

Similarly, we can plot the runtime that is acquired by numerically solving the transcendental equations, and compare it with the one obtained using the exact search Hamiltonian \eqref{eq:H}. This is shown in Fig.~\ref{fig:time_k_M1000}, and although there is a slight discrepancy of roughly 5 time units for some values of $k$, this is too small to significantly affect the algorithm because the success probability has a rather wide peak.

\section{Large, XL, and XXL Weights}

\begin{figure}
\begin{center}
	\subfloat[]{
		\label{fig:diagram_H0_wL}
		\includegraphics{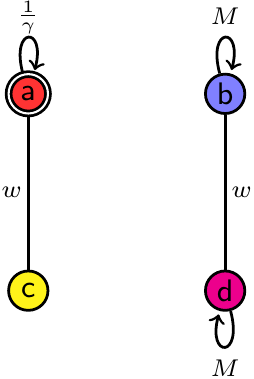}
	} \quad \quad \quad \quad
	\subfloat[]{
		\label{fig:diagram_H01_wL}
		\includegraphics{diagram_H01_wL}
	}
	\caption{\label{fig:diagram_wL} Apart from a factor of $-\gamma$, (a) the leading-order Hamiltonian with large, extra large, or extra extra large weights, and (b) with a perturbation.}
\end{center}
\end{figure}

In this section, we examine when $w$ scales larger than $\sqrt{M}$, which encompasses the remaining large, extra large, and extra extra large cases. That is, $w = \omega(\sqrt{M})$. We take the leading-order search Hamiltonian \eqref{eq:H} to be
\[ H^{(0)} =  -\gamma \begin{pmatrix}
	\frac{1}{\gamma} & 0 & w & 0 \\
	0 & M & 0 & w \\
	w & 0 & 0 & 0 \\
	0 & w & 0 & M \\
\end{pmatrix}. \]
This is visualized in Fig.~\ref{fig:diagram_H0_wL}. Its eigenvectors and corresponding eigenvalues are
\begin{gather*}
	u = \frac{1 + \sqrt{1+4w^2\gamma^2}}{2 w \gamma} \ket{a} + \ket{c}, \quad \frac{-1}{2} \left( 1 + \sqrt{1+4w^2\gamma^2} \right) \\
	\frac{1 - \sqrt{1+4w^2\gamma^2}}{2 w \gamma} \ket{a} + \ket{c}, \quad \frac{-1}{2} \left( 1 - \sqrt{1+4w^2\gamma^2} \right) \\
	\ket{v} = \frac{1}{\sqrt{2}} \left( \ket{b} + \ket{d} \right), \quad -\gamma(M+w) \\
	\frac{1}{\sqrt{2}} \left( -\ket{b} + \ket{d} \right), \quad -\gamma(M-w).
\end{gather*}
Note that the first two eigenvectors are unnormalized. Setting the first and third eigenvalues equal to each other, we get the critical $\gamma$:
\[ \gamma_c = \frac{M+w}{M(M+2w)}. \]
With this value of $\gamma$, the first eigenvector $u$ is exactly
\[ u = \left( \frac{M}{w} + 1 \right) \ket{a} + \ket{c}, \]
and normalizing it yields
\[ \ket{u} = \frac{w}{\sqrt{(M+w)^2 + w^2}} \left( \frac{M+w}{w} \ket{a} + \ket{c} \right). \]

Although we derived this $\gamma_c$ for the large, XL, and XXL weight cases, it also works for small and medium weights. Taylor expanding it for $w = o(M)$, we get
\[ \frac{M+w}{M(M+2w)} = \frac{1}{M} - \frac{w}{M^2} + \frac{2w^2}{M^3} + O\left( \frac{w^3}{M^4} \right), \]
This successfully encompasses $1/M$ for small weights and the medium weight correction of $-w/M^2$, which it must for continuity of $\gamma_c$ as the weight changes. Thus $(M+w)/M(M+2w)$ can be used as the critical $\gamma$ for all cases of weights, such as for computing Fig.~\ref{fig:prob_time_M1000}.

With the perturbation $H^{(1)}$, which restores terms of $\Theta(\sqrt{M})$, the Hamiltonian becomes
\[ H^{(0)} + H^{(1)} =  -\gamma \begin{pmatrix}
	\frac{1}{\gamma} & \sqrt{M} & w & 0 \\
	\sqrt{M} & M & 0 & w \\
	w & 0 & 0 & \sqrt{M} \\
	0 & w & \sqrt{M} & M \\
\end{pmatrix}. \]
This is visualized in Fig.~\ref{fig:diagram_H01_wL}. From the restored edges, we see that probability can flow between all four types of vertices. So there is the possibility that the success probability can be greater than the unweighted (or small weight) value of $1/2$. With the perturbation, two of the eigenvectors are $\alpha_u \ket{u} + \alpha_v \ket{v}$, where
\[ \begin{pmatrix}
	H_{uu} & H_{uv} \\
	H_{vu} & H_{vv} \\
\end{pmatrix} \begin{pmatrix}
	\alpha_u \\
	\alpha_v \\
\end{pmatrix} = E \begin{pmatrix}
	\alpha_u \\
	\alpha_v \\
\end{pmatrix}, \]
where $H_{uv} = \bra{u} H^{(0)}+H^{(1)} \ket{v}$, etc. Evaluating these matrix components, we get
\[ \begin{pmatrix}
	\frac{-(M+w)^2}{M(M+2w)} & \frac{-(M+w)}{\sqrt{2M}\sqrt{(M+w)^2 + w^2}} \\
	\frac{-(M+w)}{\sqrt{2M}\sqrt{(M+w)^2 + w^2}} & \frac{-(M+w)^2}{M(M+2w)} \\
\end{pmatrix} \!\!\! \begin{pmatrix}
	\alpha_u \\
	\alpha_v \\
\end{pmatrix} \!\! = \!  E \! \begin{pmatrix}
	\alpha_u \\
	\alpha_v \\
\end{pmatrix} \!\!. \]
Solving this, we get perturbed eigenstates
\[ \frac{1}{\sqrt{2}} \left( \ket{v} \pm \ket{u} \right) \]
with corresponding eigenvalues
\[ -1 - \frac{w}{2M} + \frac{w}{2(M+2w)} \mp \frac{M+w}{\sqrt{2M}\sqrt{(M+w)^2+w^2}}. \]
Thus the system evolves from $\ket{s} \approx \ket{v}$ to $\ket{u}$ in time $\pi/\Delta E$ \cite{CG2004,Wong10}, or
\[ t_* = \frac{\pi}{\sqrt{2}} \frac{\sqrt{M}\sqrt{(M+w)^2+w^2}}{M+w}. \]
To find the success probability, we simply find how much of $\ket{u}$ comes from $\ket{a}$, i.e.,
\[ p_* = |\braket{a}{u}|^2 = \frac{(M+w)^2}{(M+w)^2 + w^2}. \]
These results are consistent with Fig.~\ref{fig:prob_time_M1000_w100-20000}. For example, when $M = 1000$ and $w = 100$, our analytical formulas yield $t_* \approx 70.538$ and $p_* \approx 0.99$, which are consistent with the figure. Similarly, when $w = 3000$, our formulas yield $t_* \approx 87.810$ and $p_* = 0.64$, which are also consistent with the figure.

Note that the quantity $p_* = |\braket{a}{u}|^2$ is only the probability of measuring the particle at the marked vertex $a$. As noted in Section II, however, the final state $\ket{u}$ is a linear combination of $\ket{a}$ and $\ket{c}$. So measuring the randomly walking quantum particle in this state, we find the particle at vertex $a$ with probability $p_*$ and at vertex $c$ with probability $1-p_*$. Whether the particle is at $\ket{a}$ or $\ket{c}$ can be determined by a single oracle query, which is negligible compared to the $\Theta(\sqrt{M})$ queries of the algorithm. Then from Fig.~\ref{fig:linkedcomplete}, if the result is $c$, the marked vertex $a$ can be directly inferred as the neighbor of $c$ in the other clique. Thus the true success probability of the algorithm is $1$, not $p_*$. So there is no need to repeat the algorithm, and the expected runtime is simply a single runtime.

With large or extra extra large weights, we can further simplify the runtime and probability $p_*$ for large $M$. With large weights, $w = o(M)$, so the runtime is asymptotically
\[ t_* = \frac{\pi}{\sqrt{2}} \sqrt{M}, \]
with corresponding asymptotic probability
\[ p_* = 1. \]
In other words, the system evolves from $\ket{s}$ to $\ket{a}$, so there is no need to infer $\ket{a}$ from $\ket{c}$.

With extra extra large weights, $w = \omega(M)$, so the runtime is asymptotically
\[ t_* = \pi \sqrt{M}, \]
and the success probability is asymptotically
\[ p_* = \frac{1}{2}. \]
So the final state $\ket{u}$ is half in $\ket{a}$ and half in $\ket{c}$. Again, vertex $a$ can be inferred from vertex $c$, so the true success probability is $1$.

These results are also summarized in Table~\ref{table:summary}, and they complete our analysis of all five cases of weights. Using the table, it is easy to identify the behavior of the algorithm as the weights increase: If the graph is unweighted (i.e., $w = 1$) or has small weights, the runtime of a single iteration of the algorithm is $\pi\sqrt{M}/2$ with a success probability is $1/2$, which yields an expected runtime with classical repetitions of $\pi\sqrt{M}$. As the weights increase through the medium case to the large case, the success probability doubles to $1$ and while the runtime increases by a factor of $\sqrt{2}$, so the expected runtime achieves an overall improvement by a factor of $\sqrt{2}$. If the weights are further increased through the XL case to the XXL case, however, the success probability effectively remains at $1$ while the runtime is lengthened by a factor of $\sqrt{2}$, eliminating the benefits of the probability boost and returning the expected runtime to $\pi\sqrt{M}$. Thus the weights can be engineered to improve the search, but they must be chosen judiciously to not be too weak nor too strong. In particular, the large weight case is optimal.

Note that a constant-factor improvement in the expected runtime, which we achieved, is the greatest that one can expect for our problem given the optimality of Grover's algorithm---one cannot search faster than $O(\sqrt{M})$ \cite{BBBV1997}.


\section{Energy}

We end by commenting on the energy usage of the algorithm, which can be quantified by the operator norm of the search Hamiltonian \eqref{eq:H}. Recall $H$ is composed of a quantum walk term $-\gamma A$ (since $D$ can be dropped) and an oracle term $\ketbra{a}{a}$. The adjacency matrix has operator norm $M+w-1$. Multiplying by the critical $\gamma$, the quantum walk term has operator norm
\[ \frac{M+w}{M(M+2w)} (M+w-1). \]
When $w = o(M)$, which includes the small, medium, and large weight cases, then this is asymptotically $\Theta(1)$. The oracle term also has operator norm $1$, so for these weights, the search Hamiltonian has operator norm $\Theta(1)$. Since the large weight case is the one for which the algorithm is optimal, this improvement does not asymptotically require more energy.

This may be expected since the success probability is only improved by a constant factor. In previous work \cite{Wong16}, however, a speedup in runtime scaling was also achieved without increasing the asymptotic energy usage of the algorithm. So one might expect that, for other problems, larger improvements in the success probability are possible within energy constraints.


\section{Conclusion}

In physical systems, quantum walks may occur on weighted graphs. Some previous work has explored how to engineer the weights to improve state transfer, but here we showed that the success probability of an algorithm can also be improved by weighing edges correctly. For searching on the linked complete graphs, increasing the weights of the edges between the two complete graphs can boost the success probability from $1/2$ to $1$. If one further increases the weights, however, the success probability stays at $1$ while the runtime is worsened, eliminating the benefit of the probability doubling. So there is an intermediate zone where the algorithm is optimized. Thus improving the algorithm is not a trivial procedure of making weights as strong as possible---some engineering is involved.

Further research includes investigating other algorithms based on quantum walks to see if they can similarly be improved by weighing the edges.


\begin{acknowledgments}
	T.W.~was supported by the European Union Seventh Framework Programme (FP7/2007-2013) under the QALGO (Grant Agreement No.~600700) project, and the ERC Advanced Grant MQC. P.P.~was supported by CNPq CSF/BJT grant reference 400216/2014-0.
\end{acknowledgments}


\bibliography{refs}

\begin{thebibliography}{28}%
\makeatletter
\providecommand \@ifxundefined [1]{%
 \@ifx{#1\undefined}
}%
\providecommand \@ifnum [1]{%
 \ifnum #1\expandafter \@firstoftwo
 \else \expandafter \@secondoftwo
 \fi
}%
\providecommand \@ifx [1]{%
 \ifx #1\expandafter \@firstoftwo
 \else \expandafter \@secondoftwo
 \fi
}%
\providecommand \natexlab [1]{#1}%
\providecommand \enquote  [1]{``#1''}%
\providecommand \bibnamefont  [1]{#1}%
\providecommand \bibfnamefont [1]{#1}%
\providecommand \citenamefont [1]{#1}%
\providecommand \href@noop [0]{\@secondoftwo}%
\providecommand \href [0]{\begingroup \@sanitize@url \@href}%
\providecommand \@href[1]{\@@startlink{#1}\@@href}%
\providecommand \@@href[1]{\endgroup#1\@@endlink}%
\providecommand \@sanitize@url [0]{\catcode `\\12\catcode `\$12\catcode
  `\&12\catcode `\#12\catcode `\^12\catcode `\_12\catcode `\%12\relax}%
\providecommand \@@startlink[1]{}%
\providecommand \@@endlink[0]{}%
\providecommand \url  [0]{\begingroup\@sanitize@url \@url }%
\providecommand \@url [1]{\endgroup\@href {#1}{\urlprefix }}%
\providecommand \urlprefix  [0]{URL }%
\providecommand \Eprint [0]{\href }%
\providecommand \doibase [0]{http://dx.doi.org/}%
\providecommand \selectlanguage [0]{\@gobble}%
\providecommand \bibinfo  [0]{\@secondoftwo}%
\providecommand \bibfield  [0]{\@secondoftwo}%
\providecommand \translation [1]{[#1]}%
\providecommand \BibitemOpen [0]{}%
\providecommand \bibitemStop [0]{}%
\providecommand \bibitemNoStop [0]{.\EOS\space}%
\providecommand \EOS [0]{\spacefactor3000\relax}%
\providecommand \BibitemShut  [1]{\csname bibitem#1\endcsname}%
\let\auto@bib@innerbib\@empty
\bibitem [{\citenamefont {Farhi}\ and\ \citenamefont {Gutmann}(1998)}]{FG1998}%
  \BibitemOpen
  \bibfield  {author} {\bibinfo {author} {\bibfnamefont {Edward}\ \bibnamefont
  {Farhi}}\ and\ \bibinfo {author} {\bibfnamefont {Sam}\ \bibnamefont
  {Gutmann}},\ }\bibfield  {title} {\enquote {\bibinfo {title} {Quantum
  computation and decision trees},}\ }\href {\doibase 10.1103/PhysRevA.58.915}
  {\bibfield  {journal} {\bibinfo  {journal} {Phys. Rev. A}\ }\textbf {\bibinfo
  {volume} {58}},\ \bibinfo {pages} {915--928} (\bibinfo {year}
  {1998})}\BibitemShut {NoStop}%
\bibitem [{\citenamefont {Farhi}\ \emph {et~al.}(2008)\citenamefont {Farhi},
  \citenamefont {Goldstone},\ and\ \citenamefont {Gutmann}}]{FGG2008}%
  \BibitemOpen
  \bibfield  {author} {\bibinfo {author} {\bibfnamefont {Edward}\ \bibnamefont
  {Farhi}}, \bibinfo {author} {\bibfnamefont {Jeffrey}\ \bibnamefont
  {Goldstone}}, \ and\ \bibinfo {author} {\bibfnamefont {Sam}\ \bibnamefont
  {Gutmann}},\ }\bibfield  {title} {\enquote {\bibinfo {title} {A quantum
  algorithm for the {H}amiltonian {NAND} tree},}\ }\href {\doibase
  10.4086/toc.2008.v004a008} {\bibfield  {journal} {\bibinfo  {journal} {Theory
  Comput.}\ }\textbf {\bibinfo {volume} {4}},\ \bibinfo {pages} {169--190}
  (\bibinfo {year} {2008})}\BibitemShut {NoStop}%
\bibitem [{\citenamefont {Childs}\ and\ \citenamefont
  {Goldstone}(2004)}]{CG2004}%
  \BibitemOpen
  \bibfield  {author} {\bibinfo {author} {\bibfnamefont {Andrew~M.}\
  \bibnamefont {Childs}}\ and\ \bibinfo {author} {\bibfnamefont {Jeffrey}\
  \bibnamefont {Goldstone}},\ }\bibfield  {title} {\enquote {\bibinfo {title}
  {Spatial search by quantum walk},}\ }\href {\doibase
  10.1103/PhysRevA.70.022314} {\bibfield  {journal} {\bibinfo  {journal} {Phys.
  Rev. A}\ }\textbf {\bibinfo {volume} {70}},\ \bibinfo {pages} {022314}
  (\bibinfo {year} {2004})}\BibitemShut {NoStop}%
\bibitem [{\citenamefont {Ambainis}(2003)}]{Ambainis2003}%
  \BibitemOpen
  \bibfield  {author} {\bibinfo {author} {\bibfnamefont {Andris}\ \bibnamefont
  {Ambainis}},\ }\bibfield  {title} {\enquote {\bibinfo {title} {Quantum walks
  and their algorithmic applications},}\ }\href {\doibase
  10.1142/S0219749903000383} {\bibfield  {journal} {\bibinfo  {journal} {Int.
  J. Quantum Inf.}\ }\textbf {\bibinfo {volume} {01}},\ \bibinfo {pages}
  {507--518} (\bibinfo {year} {2003})}\BibitemShut {NoStop}%
\bibitem [{\citenamefont {Childs}(2010)}]{Childs2010}%
  \BibitemOpen
  \bibfield  {author} {\bibinfo {author} {\bibfnamefont {Andrew~M.}\
  \bibnamefont {Childs}},\ }\bibfield  {title} {\enquote {\bibinfo {title} {On
  the relationship between continuous- and discrete-time quantum walk},}\
  }\href {\doibase 10.1007/s00220-009-0930-1} {\bibfield  {journal} {\bibinfo
  {journal} {Commun. Math. Phys.}\ }\textbf {\bibinfo {volume} {294}},\
  \bibinfo {pages} {581--603} (\bibinfo {year} {2010})}\BibitemShut {NoStop}%
\bibitem [{\citenamefont {Childs}\ \emph {et~al.}(2003)\citenamefont {Childs},
  \citenamefont {Cleve}, \citenamefont {Deotto}, \citenamefont {Farhi},
  \citenamefont {Gutmann},\ and\ \citenamefont {Spielman}}]{CCDFGS2003}%
  \BibitemOpen
  \bibfield  {author} {\bibinfo {author} {\bibfnamefont {Andrew~M.}\
  \bibnamefont {Childs}}, \bibinfo {author} {\bibfnamefont {Richard}\
  \bibnamefont {Cleve}}, \bibinfo {author} {\bibfnamefont {Enrico}\
  \bibnamefont {Deotto}}, \bibinfo {author} {\bibfnamefont {Edward}\
  \bibnamefont {Farhi}}, \bibinfo {author} {\bibfnamefont {Sam}\ \bibnamefont
  {Gutmann}}, \ and\ \bibinfo {author} {\bibfnamefont {Daniel~A.}\ \bibnamefont
  {Spielman}},\ }\bibfield  {title} {\enquote {\bibinfo {title} {Exponential
  algorithmic speedup by a quantum walk},}\ }in\ \href {\doibase
  10.1145/780542.780552} {\emph {\bibinfo {booktitle} {Proceedings of the 35th
  Annual ACM Symposium on Theory of Computing}}},\ \bibinfo {series and number}
  {STOC '03}\ (\bibinfo  {publisher} {ACM},\ \bibinfo {address} {New York, NY,
  USA},\ \bibinfo {year} {2003})\ pp.\ \bibinfo {pages} {59--68}\BibitemShut
  {NoStop}%
\bibitem [{\citenamefont {Mohseni}\ \emph {et~al.}(2008)\citenamefont
  {Mohseni}, \citenamefont {Rebentrost}, \citenamefont {Lloyd},\ and\
  \citenamefont {Aspuru-Guzik}}]{Mohseni2008}%
  \BibitemOpen
  \bibfield  {author} {\bibinfo {author} {\bibfnamefont {Masoud}\ \bibnamefont
  {Mohseni}}, \bibinfo {author} {\bibfnamefont {Patrick}\ \bibnamefont
  {Rebentrost}}, \bibinfo {author} {\bibfnamefont {Seth}\ \bibnamefont
  {Lloyd}}, \ and\ \bibinfo {author} {\bibfnamefont {Al\'an}\ \bibnamefont
  {Aspuru-Guzik}},\ }\bibfield  {title} {\enquote {\bibinfo {title}
  {Environment-assisted quantum walks in photosynthetic energy transfer},}\
  }\href {\doibase 10.1063/1.3002335} {\bibfield  {journal} {\bibinfo
  {journal} {J. Chem. Phys.}\ }\textbf {\bibinfo {volume} {129}},\ \bibinfo
  {pages} {174106} (\bibinfo {year} {2008})}\BibitemShut {NoStop}%
\bibitem [{\citenamefont {Carlson}\ \emph {et~al.}(2007)\citenamefont
  {Carlson}, \citenamefont {Ford}, \citenamefont {Harris}, \citenamefont
  {Rosen}, \citenamefont {Tamon},\ and\ \citenamefont {Wrobel}}]{CFHRTW2007}%
  \BibitemOpen
  \bibfield  {author} {\bibinfo {author} {\bibfnamefont {William}\ \bibnamefont
  {Carlson}}, \bibinfo {author} {\bibfnamefont {Allison}\ \bibnamefont {Ford}},
  \bibinfo {author} {\bibfnamefont {Elizabeth}\ \bibnamefont {Harris}},
  \bibinfo {author} {\bibfnamefont {Julian}\ \bibnamefont {Rosen}}, \bibinfo
  {author} {\bibfnamefont {Christino}\ \bibnamefont {Tamon}}, \ and\ \bibinfo
  {author} {\bibfnamefont {Kathleen}\ \bibnamefont {Wrobel}},\ }\bibfield
  {title} {\enquote {\bibinfo {title} {Universal mixing of quantum walk on
  graphs},}\ }\href@noop {} {\bibfield  {journal} {\bibinfo  {journal} {Quantum
  Inf. Comput.}\ }\textbf {\bibinfo {volume} {7}},\ \bibinfo {pages} {738--751}
  (\bibinfo {year} {2007})}\BibitemShut {NoStop}%
\bibitem [{\citenamefont {Christandl}\ \emph {et~al.}(2004)\citenamefont
  {Christandl}, \citenamefont {Datta}, \citenamefont {Ekert},\ and\
  \citenamefont {Landahl}}]{CDEL2004}%
  \BibitemOpen
  \bibfield  {author} {\bibinfo {author} {\bibfnamefont {Matthias}\
  \bibnamefont {Christandl}}, \bibinfo {author} {\bibfnamefont {Nilanjana}\
  \bibnamefont {Datta}}, \bibinfo {author} {\bibfnamefont {Artur}\ \bibnamefont
  {Ekert}}, \ and\ \bibinfo {author} {\bibfnamefont {Andrew~J.}\ \bibnamefont
  {Landahl}},\ }\bibfield  {title} {\enquote {\bibinfo {title} {Perfect state
  transfer in quantum spin networks},}\ }\href {\doibase
  10.1103/PhysRevLett.92.187902} {\bibfield  {journal} {\bibinfo  {journal}
  {Phys. Rev. Lett.}\ }\textbf {\bibinfo {volume} {92}},\ \bibinfo {pages}
  {187902} (\bibinfo {year} {2004})}\BibitemShut {NoStop}%
\bibitem [{\citenamefont {Zimbor\'as}\ \emph {et~al.}(2013)\citenamefont
  {Zimbor\'as}, \citenamefont {Faccin}, \citenamefont {K\'ad\'ar},
  \citenamefont {Whitfield}, \citenamefont {Lanyon},\ and\ \citenamefont
  {Biamonte}}]{Zimboras2013}%
  \BibitemOpen
  \bibfield  {author} {\bibinfo {author} {\bibfnamefont {Zolt\'an}\
  \bibnamefont {Zimbor\'as}}, \bibinfo {author} {\bibfnamefont {Mauro}\
  \bibnamefont {Faccin}}, \bibinfo {author} {\bibfnamefont {Zolt\'an}\
  \bibnamefont {K\'ad\'ar}}, \bibinfo {author} {\bibfnamefont {James~D.}\
  \bibnamefont {Whitfield}}, \bibinfo {author} {\bibfnamefont {Ben~P.}\
  \bibnamefont {Lanyon}}, \ and\ \bibinfo {author} {\bibfnamefont {Jacob}\
  \bibnamefont {Biamonte}},\ }\bibfield  {title} {\enquote {\bibinfo {title}
  {Quantum transport enhancement by time-reversal symmetry breaking},}\ }\href
  {\doibase 10.1038/srep02361} {\bibfield  {journal} {\bibinfo  {journal} {Sci.
  Rep.}\ }\textbf {\bibinfo {volume} {3}},\ \bibinfo {pages} {2361} (\bibinfo
  {year} {2013})}\BibitemShut {NoStop}%
\bibitem [{\citenamefont {Lu}\ \emph {et~al.}(2016)\citenamefont {Lu},
  \citenamefont {Biamonte}, \citenamefont {Li}, \citenamefont {Li},
  \citenamefont {Johnson}, \citenamefont {Bergholm}, \citenamefont {Faccin},
  \citenamefont {Zimbor\'as}, \citenamefont {Laflamme}, \citenamefont {Baugh},\
  and\ \citenamefont {Lloyd}}]{Lu2016}%
  \BibitemOpen
  \bibfield  {author} {\bibinfo {author} {\bibfnamefont {Dawei}\ \bibnamefont
  {Lu}}, \bibinfo {author} {\bibfnamefont {Jacob~D.}\ \bibnamefont {Biamonte}},
  \bibinfo {author} {\bibfnamefont {Jun}\ \bibnamefont {Li}}, \bibinfo {author}
  {\bibfnamefont {Hang}\ \bibnamefont {Li}}, \bibinfo {author} {\bibfnamefont
  {Tomi~H.}\ \bibnamefont {Johnson}}, \bibinfo {author} {\bibfnamefont {Ville}\
  \bibnamefont {Bergholm}}, \bibinfo {author} {\bibfnamefont {Mauro}\
  \bibnamefont {Faccin}}, \bibinfo {author} {\bibfnamefont {Zolt\'an}\
  \bibnamefont {Zimbor\'as}}, \bibinfo {author} {\bibfnamefont {Raymond}\
  \bibnamefont {Laflamme}}, \bibinfo {author} {\bibfnamefont {Jonathan}\
  \bibnamefont {Baugh}}, \ and\ \bibinfo {author} {\bibfnamefont {Seth}\
  \bibnamefont {Lloyd}},\ }\bibfield  {title} {\enquote {\bibinfo {title}
  {Chiral quantum walks},}\ }\href {\doibase 10.1103/PhysRevA.93.042302}
  {\bibfield  {journal} {\bibinfo  {journal} {Phys. Rev. A}\ }\textbf {\bibinfo
  {volume} {93}},\ \bibinfo {pages} {042302} (\bibinfo {year}
  {2016})}\BibitemShut {NoStop}%
\bibitem [{\citenamefont {Mochon}(2007)}]{Mochon2007}%
  \BibitemOpen
  \bibfield  {author} {\bibinfo {author} {\bibfnamefont {Carlos}\ \bibnamefont
  {Mochon}},\ }\bibfield  {title} {\enquote {\bibinfo {title} {Hamiltonian
  oracles},}\ }\href {\doibase 10.1103/PhysRevA.75.042313} {\bibfield
  {journal} {\bibinfo  {journal} {Phys. Rev. A}\ }\textbf {\bibinfo {volume}
  {75}},\ \bibinfo {pages} {042313} (\bibinfo {year} {2007})}\BibitemShut
  {NoStop}%
\bibitem [{\citenamefont {Grover}(1996)}]{Grover1996}%
  \BibitemOpen
  \bibfield  {author} {\bibinfo {author} {\bibfnamefont {Lov~K.}\ \bibnamefont
  {Grover}},\ }\bibfield  {title} {\enquote {\bibinfo {title} {A fast quantum
  mechanical algorithm for database search},}\ }in\ \href@noop {} {\emph
  {\bibinfo {booktitle} {Proceedings of the 28th Annual ACM Symposium on Theory
  of Computing}}},\ \bibinfo {series and number} {STOC '96}\ (\bibinfo
  {publisher} {ACM},\ \bibinfo {address} {New York, NY, USA},\ \bibinfo {year}
  {1996})\ pp.\ \bibinfo {pages} {212--219}\BibitemShut {NoStop}%
\bibitem [{\citenamefont {Ambainis}\ \emph {et~al.}(2005)\citenamefont
  {Ambainis}, \citenamefont {Kempe},\ and\ \citenamefont {Rivosh}}]{AKR2005}%
  \BibitemOpen
  \bibfield  {author} {\bibinfo {author} {\bibfnamefont {Andris}\ \bibnamefont
  {Ambainis}}, \bibinfo {author} {\bibfnamefont {Julia}\ \bibnamefont {Kempe}},
  \ and\ \bibinfo {author} {\bibfnamefont {Alexander}\ \bibnamefont {Rivosh}},\
  }\bibfield  {title} {\enquote {\bibinfo {title} {Coins make quantum walks
  faster},}\ }in\ \href@noop {} {\emph {\bibinfo {booktitle} {Proceedings of
  the 16th Annual ACM-SIAM Symposium on Discrete Algorithms}}},\ \bibinfo
  {series and number} {SODA '05}\ (\bibinfo  {publisher} {SIAM},\ \bibinfo
  {address} {Philadelphia, PA, USA},\ \bibinfo {year} {2005})\ pp.\ \bibinfo
  {pages} {1099--1108}\BibitemShut {NoStop}%
\bibitem [{\citenamefont {Wong}(2015{\natexlab{a}})}]{Wong10}%
  \BibitemOpen
  \bibfield  {author} {\bibinfo {author} {\bibfnamefont {Thomas~G.}\
  \bibnamefont {Wong}},\ }\bibfield  {title} {\enquote {\bibinfo {title}
  {Grover search with lackadaisical quantum walks},}\ }\href {\doibase
  10.1088/1751-8113/48/43/435304} {\bibfield  {journal} {\bibinfo  {journal}
  {J. Phys. A: Math. Theor.}\ }\textbf {\bibinfo {volume} {48}},\ \bibinfo
  {pages} {435304} (\bibinfo {year} {2015}{\natexlab{a}})}\BibitemShut
  {NoStop}%
\bibitem [{\citenamefont {Novo}\ \emph {et~al.}(2015)\citenamefont {Novo},
  \citenamefont {Chakraborty}, \citenamefont {Mohseni}, \citenamefont {Neven},\
  and\ \citenamefont {Omar}}]{Novo2015}%
  \BibitemOpen
  \bibfield  {author} {\bibinfo {author} {\bibfnamefont {Leonardo}\
  \bibnamefont {Novo}}, \bibinfo {author} {\bibfnamefont {Shantanav}\
  \bibnamefont {Chakraborty}}, \bibinfo {author} {\bibfnamefont {Masoud}\
  \bibnamefont {Mohseni}}, \bibinfo {author} {\bibfnamefont {Hartmut}\
  \bibnamefont {Neven}}, \ and\ \bibinfo {author} {\bibfnamefont {Yasser}\
  \bibnamefont {Omar}},\ }\bibfield  {title} {\enquote {\bibinfo {title}
  {Systematic dimensionality reduction for quantum walks: Optimal spatial
  search and transport on non-regular graphs},}\ }\href {\doibase
  10.1038/srep13304} {\bibfield  {journal} {\bibinfo  {journal} {Sci. Rep.}\
  }\textbf {\bibinfo {volume} {5}},\ \bibinfo {pages} {13304} (\bibinfo {year}
  {2015})}\BibitemShut {NoStop}%
\bibitem [{\citenamefont {Wong}\ \emph {et~al.}(2015)\citenamefont {Wong},
  \citenamefont {Tarrataca},\ and\ \citenamefont {Nahimov}}]{Wong19}%
  \BibitemOpen
  \bibfield  {author} {\bibinfo {author} {\bibfnamefont {Thomas~G.}\
  \bibnamefont {Wong}}, \bibinfo {author} {\bibfnamefont {Lu\'{i}s}\
  \bibnamefont {Tarrataca}}, \ and\ \bibinfo {author} {\bibfnamefont {Nikolay}\
  \bibnamefont {Nahimov}},\ }\bibfield  {title} {\enquote {\bibinfo {title}
  {Laplacian versus adjacency matrix in quantum walk search},}\ }\href@noop {}
  {\  (\bibinfo {year} {2015})},\ \Eprint
  {http://arxiv.org/abs/{a}rXiv:1512.05554 [quant-ph]} {{a}rXiv:1512.05554
  [quant-ph]} \BibitemShut {NoStop}%
\bibitem [{\citenamefont {Philipp}\ \emph {et~al.}(2016)\citenamefont
  {Philipp}, \citenamefont {Tarrataca},\ and\ \citenamefont
  {Boettcher}}]{Philipp2016}%
  \BibitemOpen
  \bibfield  {author} {\bibinfo {author} {\bibfnamefont {Pascal}\ \bibnamefont
  {Philipp}}, \bibinfo {author} {\bibfnamefont {Lu\'{\i}s}\ \bibnamefont
  {Tarrataca}}, \ and\ \bibinfo {author} {\bibfnamefont {Stefan}\ \bibnamefont
  {Boettcher}},\ }\bibfield  {title} {\enquote {\bibinfo {title}
  {Continuous-time quantum search on balanced trees},}\ }\href {\doibase
  10.1103/PhysRevA.93.032305} {\bibfield  {journal} {\bibinfo  {journal} {Phys.
  Rev. A}\ }\textbf {\bibinfo {volume} {93}},\ \bibinfo {pages} {032305}
  (\bibinfo {year} {2016})}\BibitemShut {NoStop}%
\bibitem [{\citenamefont {Chakraborty}\ \emph {et~al.}(2016)\citenamefont
  {Chakraborty}, \citenamefont {Novo}, \citenamefont {Ambainis},\ and\
  \citenamefont {Omar}}]{Chakraborty2016}%
  \BibitemOpen
  \bibfield  {author} {\bibinfo {author} {\bibfnamefont {Shantanav}\
  \bibnamefont {Chakraborty}}, \bibinfo {author} {\bibfnamefont {Leonardo}\
  \bibnamefont {Novo}}, \bibinfo {author} {\bibfnamefont {Andris}\ \bibnamefont
  {Ambainis}}, \ and\ \bibinfo {author} {\bibfnamefont {Yasser}\ \bibnamefont
  {Omar}},\ }\bibfield  {title} {\enquote {\bibinfo {title} {Spatial search by
  quantum walk is optimal for almost all graphs},}\ }\href {\doibase
  10.1103/PhysRevLett.116.100501} {\bibfield  {journal} {\bibinfo  {journal}
  {Phys. Rev. Lett.}\ }\textbf {\bibinfo {volume} {116}},\ \bibinfo {pages}
  {100501} (\bibinfo {year} {2016})}\BibitemShut {NoStop}%
\bibitem [{\citenamefont {Meyer}\ and\ \citenamefont {Wong}(2015)}]{Wong7}%
  \BibitemOpen
  \bibfield  {author} {\bibinfo {author} {\bibfnamefont {David~A.}\
  \bibnamefont {Meyer}}\ and\ \bibinfo {author} {\bibfnamefont {Thomas~G.}\
  \bibnamefont {Wong}},\ }\bibfield  {title} {\enquote {\bibinfo {title}
  {Connectivity is a poor indicator of fast quantum search},}\ }\href {\doibase
  10.1103/PhysRevLett.114.110503} {\bibfield  {journal} {\bibinfo  {journal}
  {Phys. Rev. Lett.}\ }\textbf {\bibinfo {volume} {114}},\ \bibinfo {pages}
  {110503} (\bibinfo {year} {2015})}\BibitemShut {NoStop}%
\bibitem [{\citenamefont {Wong}(2016{\natexlab{a}})}]{Wong9}%
  \BibitemOpen
  \bibfield  {author} {\bibinfo {author} {\bibfnamefont {Thomas~G.}\
  \bibnamefont {Wong}},\ }\bibfield  {title} {\enquote {\bibinfo {title}
  {Spatial search by continuous-time quantum walk with multiple marked
  vertices},}\ }\href {\doibase 10.1007/s11128-015-1239-y} {\bibfield
  {journal} {\bibinfo  {journal} {Quantum Inf. Process.}\ }\textbf {\bibinfo
  {volume} {15}},\ \bibinfo {pages} {1411--1443} (\bibinfo {year}
  {2016}{\natexlab{a}})}\BibitemShut {NoStop}%
\bibitem [{\citenamefont {Wong}\ and\ \citenamefont {Ambainis}(2015)}]{Wong11}%
  \BibitemOpen
  \bibfield  {author} {\bibinfo {author} {\bibfnamefont {Thomas~G.}\
  \bibnamefont {Wong}}\ and\ \bibinfo {author} {\bibfnamefont {Andris}\
  \bibnamefont {Ambainis}},\ }\bibfield  {title} {\enquote {\bibinfo {title}
  {Quantum search with multiple walk steps per oracle query},}\ }\href
  {\doibase 10.1103/PhysRevA.92.022338} {\bibfield  {journal} {\bibinfo
  {journal} {Phys. Rev. A}\ }\textbf {\bibinfo {volume} {92}},\ \bibinfo
  {pages} {022338} (\bibinfo {year} {2015})}\BibitemShut {NoStop}%
\bibitem [{\citenamefont {Wong}(2015{\natexlab{b}})}]{Wong16}%
  \BibitemOpen
  \bibfield  {author} {\bibinfo {author} {\bibfnamefont {Thomas~G.}\
  \bibnamefont {Wong}},\ }\bibfield  {title} {\enquote {\bibinfo {title}
  {Faster quantum walk search on a weighted graph},}\ }\href {\doibase
  10.1103/PhysRevA.92.032320} {\bibfield  {journal} {\bibinfo  {journal} {Phys.
  Rev. A}\ }\textbf {\bibinfo {volume} {92}},\ \bibinfo {pages} {032320}
  (\bibinfo {year} {2015}{\natexlab{b}})}\BibitemShut {NoStop}%
\bibitem [{\citenamefont {Wong}(2015{\natexlab{c}})}]{Wong14}%
  \BibitemOpen
  \bibfield  {author} {\bibinfo {author} {\bibfnamefont {Thomas~G.}\
  \bibnamefont {Wong}},\ }\bibfield  {title} {\enquote {\bibinfo {title}
  {Quantum walk search with time-reversal symmetry breaking},}\ }\href
  {\doibase 10.1088/1751-8113/48/40/405303} {\bibfield  {journal} {\bibinfo
  {journal} {J. Phys. A: Math. Theor.}\ }\textbf {\bibinfo {volume} {48}},\
  \bibinfo {pages} {405303} (\bibinfo {year} {2015}{\natexlab{c}})}\BibitemShut
  {NoStop}%
\bibitem [{\citenamefont {Janmark}\ \emph {et~al.}(2014)\citenamefont
  {Janmark}, \citenamefont {Meyer},\ and\ \citenamefont {Wong}}]{Wong5}%
  \BibitemOpen
  \bibfield  {author} {\bibinfo {author} {\bibfnamefont {Jonatan}\ \bibnamefont
  {Janmark}}, \bibinfo {author} {\bibfnamefont {David~A.}\ \bibnamefont
  {Meyer}}, \ and\ \bibinfo {author} {\bibfnamefont {Thomas~G.}\ \bibnamefont
  {Wong}},\ }\bibfield  {title} {\enquote {\bibinfo {title} {Global symmetry is
  unnecessary for fast quantum search},}\ }\href {\doibase
  10.1103/PhysRevLett.112.210502} {\bibfield  {journal} {\bibinfo  {journal}
  {Phys. Rev. Lett.}\ }\textbf {\bibinfo {volume} {112}},\ \bibinfo {pages}
  {210502} (\bibinfo {year} {2014})}\BibitemShut {NoStop}%
\bibitem [{\citenamefont {Wong}(2015{\natexlab{d}})}]{Wong8}%
  \BibitemOpen
  \bibfield  {author} {\bibinfo {author} {\bibfnamefont {Thomas~G.}\
  \bibnamefont {Wong}},\ }\bibfield  {title} {\enquote {\bibinfo {title}
  {Diagrammatic approach to quantum search},}\ }\href {\doibase
  10.1007/s11128-015-0959-3} {\bibfield  {journal} {\bibinfo  {journal}
  {Quantum Inf. Process.}\ }\textbf {\bibinfo {volume} {14}},\ \bibinfo {pages}
  {1767--1775} (\bibinfo {year} {2015}{\natexlab{d}})}\BibitemShut {NoStop}%
\bibitem [{\citenamefont {Wong}(2016{\natexlab{b}})}]{Wong20}%
  \BibitemOpen
  \bibfield  {author} {\bibinfo {author} {\bibfnamefont {Thomas~G.}\
  \bibnamefont {Wong}},\ }\bibfield  {title} {\enquote {\bibinfo {title}
  {Quantum walk search on {J}ohnson graphs},}\ }\href {\doibase
  10.1088/1751-8113/49/19/195303} {\bibfield  {journal} {\bibinfo  {journal}
  {J. Phys. A: Math. Theor.}\ }\textbf {\bibinfo {volume} {49}},\ \bibinfo
  {pages} {195303} (\bibinfo {year} {2016}{\natexlab{b}})}\BibitemShut
  {NoStop}%
\bibitem [{\citenamefont {Bennett}\ \emph {et~al.}(1997)\citenamefont
  {Bennett}, \citenamefont {Bernstein}, \citenamefont {Brassard},\ and\
  \citenamefont {Vazirani}}]{BBBV1997}%
  \BibitemOpen
  \bibfield  {author} {\bibinfo {author} {\bibfnamefont {Charles~H.}\
  \bibnamefont {Bennett}}, \bibinfo {author} {\bibfnamefont {Ethan}\
  \bibnamefont {Bernstein}}, \bibinfo {author} {\bibfnamefont {Gilles}\
  \bibnamefont {Brassard}}, \ and\ \bibinfo {author} {\bibfnamefont {Umesh}\
  \bibnamefont {Vazirani}},\ }\bibfield  {title} {\enquote {\bibinfo {title}
  {Strengths and weaknesses of quantum computing},}\ }\href {\doibase
  10.1137/S0097539796300933} {\bibfield  {journal} {\bibinfo  {journal} {SIAM
  Journal on Computing}\ }\textbf {\bibinfo {volume} {26}},\ \bibinfo {pages}
  {1510--1523} (\bibinfo {year} {1997})}\BibitemShut {NoStop}%
\end{thebibliography}%

\end{document}